\documentstyle[12pt,psfig]{article}
\renewcommand{\baselinestretch}{1.2}
\newcommand{\beq}{\begin{equation}}
\newcommand{\eeq}{\end{equation}}
\newcommand{\beqa}{\begin{eqnarray}}
\newcommand{\eeqa}{\end{eqnarray}}
\newcommand{\beqar}{\begin{eqnarray*}}
\newcommand{\eeqar}{\end{eqnarray*}}
\newcommand{\eg}{{\it e.g.,}\ }
\newcommand{\ie}{{\it i.e.,}\ }
\newcommand{\labell}[1]{\label{#1}}

\newcommand{\reef}[1]{(\ref{#1})}
\def\IR{{\hbox{{\rm I}\kern-.2em\hbox{\rm R}}}}
\newcommand{\bth}{{\bar\theta}}
\begin{document}

\begin{titlepage}

\begin{flushright}
DTP 99-47\\
EHU-FT/9911\\
hep-th/9906160
\end{flushright}
\vfil\vfil

\begin{center}

{\Large \bf Black diholes}

\vspace{25pt}

Roberto Emparan\footnote{roberto.emparan@durham.ac.uk}
\vfil

{\sl Department of Mathematical Sciences}\\
{\sl University of Durham, Durham DH1 3LE, UK}\\
and\\
{\sl Departamento de F{\'\i}sica Te\'orica}\\
{\sl Universidad del Pa{\'\i}s Vasco, Apdo. 644, E-48080, Bilbao, Spain}
\vfil

\end{center}

\vspace{5mm}

\begin{abstract}

\noindent 
We present and analyze exact solutions of the Einstein--Maxwell and 
Einstein--Maxwell--Dilaton equations that describe static pairs of oppositely 
charged extremal black holes, \ie black diholes. The holes are suspended in 
equilibrium in an external magnetic field, or held apart by cosmic strings. We 
comment as well on the relation of these solutions to brane--antibrane 
configurations in string and M-theory. 
\end{abstract}

\vfil\vfil\vfil
\begin{flushleft}
June 1999
\end{flushleft}
\end{titlepage}
\setcounter{footnote}{0}

\newpage
\renewcommand{\baselinestretch}{1.2}  

Exact solutions of General Relativity describing multiple black holes are few
and far between.  Indeed one would expect such configurations to have in
general a very complicated structure.  Luckily, there exist some simple
solutions exhibiting remarkable properties.  For instance, the
Majumdar--Papapetrou solutions \cite{mp} describe an arbitrary number of static
extremal charged black holes, all with charges of the same sign.  Equilibrium
is possible due to the cancellation of the gravitational attraction against
electric or magnetic repulsion.  In other cases, such as in the
multi-Schwarzschild solution in \cite{ik}, the masses are arranged in a linear
configuration, and since the gravitational attraction between them is
unbalanced, conical singularities arise along the symmetry axis.  Other
solutions describe black holes in relative motion, such as in the cosmological
multi-black hole solutions of \cite{katr}, or in the C and Ernst metrics
\cite{kin,ernst}, where two black holes accelerate apart.  In this paper we
want to report on a different class of solutions, which describe two static
extremal magnetic black holes, this time with charges of opposite signs.  The
configuration therefore possesses a magnetic dipole moment, and can be
appropriately called a {\it dihole}.  In order to maintain the black holes in
static equilibrium an external force has to be provided.  This will appear in
the form of a magnetic field aligned with the dihole. Otherwise, conical 
singularities (which may be interpreted as cosmic strings) will appear in the 
solution.

The diholes we will exhibit are solutions of Einstein--Maxwell theory, possibly
coupled to a dilaton.  The latter case includes in particular Kaluza--Klein
theory, for which the dihole consists of a monopole--antimonopole pair
described previously in \cite{dggh2}.  Dipole configurations have become of
recent interest also within the broader context of string and M-theory, as
describing brane--antibrane configurations \cite{sen}.  Near the end, we will
explain that much of what we will describe below has direct relevance in that
context.  Other recent papers studying self--gravitating dipole solutions
in string theory include
\cite{bert}.

The starting point in the construction of the new dihole solutions will be 
certain exact solutions that are known to carry magnetic dipole moment
\cite{bonnor,davged}.  We will see later that, even in the absence of an 
external magnetic field, these solutions admit an interpretation as dihole
configurations, although conical singularities will be present in general.  
For simplicity of presentation, and also because it is presumably the most
important case, we will study first the dihole in Einstein--Maxwell theory.
The extension to dilaton theories will then be a rather straightforward task.

Several years ago \cite{bonnor} Bonnor constructed a solution of 
Einstein--Maxwell theory describing a magnetic dipole, with metric
\beq
ds^2=\left( 1-{2 M r\over\Sigma}\right)^2 \left[-dt^2 + {\Sigma^4\over 
\left(\Delta 
+(M^2 +a^2)\sin^2\theta\right)^3} \left({dr^2\over \Delta}+d\theta^2 
\right)\right]+{\Delta \sin^2\theta\over \left( 1-{2 M 
r\over\Sigma}\right)^2}d\varphi^2 \ ,
\labell{bonnorm}
\eeq
and gauge potential
\beq
A={2 a M r\sin^2\theta \over \Delta+a^2\sin^2\theta}\; d\varphi\ ,
\labell{bonnora}
\eeq
with
\beqa
\Delta&=&r^2 -2 M r -a^2\ ,\nonumber\\
\Sigma&=&r^2-a^2\cos^2\theta\ .
\labell{defdesi}
\eeqa
The solution is asymptotically flat, static and axially 
symmetric. From the asymptotic behavior of $g_{tt}$ it is easy to deduce that 
the mass 
of the solution is $2 M$. The magnetic dipole moment of the solution, 
$\mu=2Ma$, becomes evident by examining the asymptotic form of the potential 
\reef{bonnora}. Changing the sign of $a$ amounts simply to reversing the 
orientation of the dipole, so we will consider, without loss of generality, 
$a\geq 0$. For $M=0$ the solution is exactly flat. It 
was noticed in \cite{bonnor} that singularities occur at $r=r_+=M+\sqrt{M^2 
+a^2}$, 
where $\Delta$ vanishes. Our aim is, first, to study the structure 
of the singularity at $r=r_+$, and show that it can be removed by the 
introduction of an external magnetic field. Then, having the new solution with 
an external magnetic field, we will argue that at the 
endpoints of the $r=r_+$ line, \ie at $(r=r_+, \theta=0)$ and $(r=r_+, 
\theta=\pi)$ lie two oppositely charged extremal Reissner-Nordstr\"{o}m black 
holes, \ie the solution describes a {\it dihole}. It will be clear then that 
the role played by the external magnetic field is to balance the attraction, 
gravitational and magnetic, between the black holes.

\begin{figure}
\hskip1cm
\psfig{figure=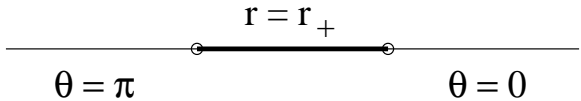}%
\hskip2cm
\caption{Axis of symmetry of the dipole solution. The lines $\theta=0,\pi$ run 
along $r_+ < r < \infty$. The segment $r=r_+$ is parametrized by $\theta$ with 
range $0<\theta < \pi$.}
\label{fig:axis}
\end{figure}

Let us then study the locus of $r=r_+$.  Crucially,
observe that the axial Killing vector $\partial_\varphi$ vanishes there.  This
means that $r=r_+$ is to be thought of as part of the symmetry axis of the
solution.  We are used to thinking of the lines $\theta=0,\pi$ as forming the
axis of symmetry.  However, in the present situation the endpoints of these two
semi-axes do not come to join at a common point.  Rather, the axis of symmetry
is completed by the segment $r=r_+$.  As $\theta$ varies from $0$ to $\pi$ we
move along this segment from one endpoint to the other, see
Fig.~\ref{fig:axis}.

The obvious thing to study now is whether conical singularities appear
on the different portions of the symmetry axis. If $\cal C$ is the
proper length of a circumference around the axis, and $\cal R$ is its
proper radius, then the presence of a conical deficit $\delta$ means
that $(d{\cal C}/d{\cal R})|_{{\cal R}\rightarrow 0}=2\pi-\delta$.
Take
$\varphi$ to be
periodically identified with period $\triangle\varphi$. Then the conical 
deficit along the axes $\theta=0,\pi$ is
\beqa
\delta_{(0,\pi)}=2\pi-\left|{\triangle\varphi\;
d\sqrt{g_{\varphi\varphi}}\over
\sqrt{g_{\theta\theta}}\;d\theta}\right|_{\theta=0,\phi}=2\pi-
\triangle\varphi
\labell{coneout}
\eeqa
and therefore would vanish with the standard choice $\triangle\varphi=2\pi$. 
However, 
the conical deficit along the line $r=r_+$,
\beq
\delta_{(r_+)}=2\pi-\left|{\triangle\varphi\;
d\sqrt{g_{\varphi\varphi}}\over
\sqrt{g_{rr}}\;dr}\right|_{r=r_+}=
 2\pi-\left(1+{M^2\over a^2}\right)^2 \triangle\varphi
\labell{conein}
\eeq
does not cancel with that same choice for $\triangle\varphi$. In fact 
$\triangle\varphi=2\pi$ gives a conical excess. This is, there is a strut along 
the segment $r=r_+$. We can see the physical origin of the strut as providing 
the internal stress (pressure) needed to counterbalance the attraction between 
the poles.

Instead of eliminating the conical defect outside the dipole, the period 
$\triangle\varphi$ can be chosen to cancel the singularity along $r=r_+$.
With such a choice one finds a conical deficit running along the axes 
$\theta=0,\pi$, from the endpoints of the dipole to infinity. We 
can view such defects as 
``cosmic strings," with tension 
\beq
T={\delta_{(0,\pi)}\over 8\pi}={1\over 4}\left[ 1-\left({a^2\over M^2 
+a^2}\right)^2\right]\ .
\labell{tension}\eeq
The dipole is then suspended by open cosmic strings that pull from its 
endpoints. The line $r=r_+$, $0<\theta<\pi$, joining these is now completely 
non-singular. 

Although the proper length of the segment $r=r_+$, $0\leq \theta \leq \pi$ is
infinite, the parameter $a$ gives, in a sense, an indication of the separation
between the poles.  For large values of $a$ the force required to keep the
dipole static becomes $T\rightarrow {M^2 \over 2 a^2}$, which decreases as
$a^{-2}$ as expected from a Newtonian approximation to the attraction between
poles.  Notice however that in the limit $a\rightarrow 0$ the tension tends to
a finite limit $T\rightarrow 1/4$.  In this limit the magnetic dipole moment of
the solution vanishes but nevertheless one does not recover the Schwarzschild
solution.  Rather, a nakedly singular solution appears, with higher
mass--multipoles \cite{bonnor}.

The recourse to cosmic strings to account for the conical singularities of the 
metric might appear as a rather {\it ad hoc} prescription. From a physical 
standpoint 
it appears that an external magnetic field aligned with the dipole should be 
able to provide the 
necessary force to balance the attraction between the poles, by pulling apart 
the dipole endpoints. By adequately tuning the magnetic field, the stresses 
along the axis should be made to disappear. 

It is indeed possible to introduce such a magnetic field by means of a Harrison
transformation \cite{harrison} on the solution.  In doing so we proceed in a
manner entirely analogous to Ernst's elimination of the conical singularities
of the C-metric \cite{ernst}.  The Harrison transformation of Einstein-Maxwell
theory takes an axisymmetric solution to another solution containing a magnetic
field that asymptotes to the Melvin magnetic universe \cite{melvin}. This is a 
flux tube that provides the best possible approximation to a uniform magnetic 
field in General Relativity.  

For an axisymmetric solution of the Einstein-Maxwell theory with 
$g_{i\varphi}=A_i=0$ for 
$i\neq \varphi$, the Harrison transformation acts as
\beqa
g'_{ij}&=&\lambda^2 g_{ij}\quad {\rm for}\;\; i,j\neq\varphi\ ,\qquad 
g'_{\varphi\varphi}=\lambda^{-2} g_{\varphi\varphi}\ ,\nonumber\\
A'_\varphi&=&{2\over \lambda B}\left(1+{B A_\varphi\over 2}\right) +k
\ ,\nonumber\\
\lambda &=& \left(1+{B A_\varphi\over 2}\right)^2+{B^2\over 4} 
g_{\varphi\varphi}\ ,
\eeqa
where $k$ is an arbitrary constant that can be chosen so as to remove Dirac 
strings.

We apply now this transformation to the Bonnor solution, eqs.~\reef{bonnorm} 
and 
\reef{bonnora}, and obtain, 
after some algebra, and choosing $k=-2/B$,
\beq
ds^2=\Lambda^2 \left[-dt^2 + {\Sigma^4\over \left(\Delta +(M^2 
+a^2)\sin^2\theta\right)^3} \left({dr^2\over \Delta}+d\theta^2 
\right)\right]+{\Delta 
\sin^2\theta \over \Lambda^2}d\varphi^2 \ ,
\labell{diholem}
\eeq
and
\beq
A_\varphi=-{2 M r a+{1\over 2} B[(r^2-a^2)^2+\Delta a^2\sin^2\theta] \over 
\Lambda\Sigma}\sin^2\theta \ ,
\labell{diholea}
\eeq
where $\Delta$ and $\Sigma$ are as in \reef{defdesi} and
\beq
\Lambda={ \Delta +a^2\sin^2\theta +2 B M r a \sin^2\theta+{1\over 4} B^2 
\sin^2\theta[(r^2-a^2)^2+\Delta a^2\sin^2\theta] \over \Sigma} \ .
\eeq
It is straightforward to see that as $r\rightarrow\infty$ the solution 
approaches the same limit as the Melvin universe with axial magnetic field $B$. 
Let us investigate now the conical structure along the symmetry axis. Along the 
outer semi-axes, $\theta=0,\pi$, we find the same value for the conical defect 
as in \reef{coneout}, so, in order to set $\delta_{(0,\pi)}=0$, we will choose 
$\triangle\varphi=2\pi$. On the other hand, along the inner segment of the 
axis, $r=r_+$, we find now
\beq
\delta_{(r_+)}=2\pi-\left(1+{M^2\over a^2}\right)^2 \left(1+{B M r_+\over 
a}\right)^{-4}\triangle\varphi \ ,
\eeq
and with the choice $\triangle\varphi=2\pi$ we see that the conical defect can 
be 
cancelled if the magnetic field is chosen to be
\beq
B={\pm\sqrt{M^2 +a^2}-a \over M r_+}=\pm{2 M\over (r_+ \pm a)^2}\ .
\labell{bvalue}
\eeq
There are two branches of solutions, one with $B>0$ and another one with $B<0$ 
(recall that we are taking $a\geq 0$). For the first branch, in the limit 
$a\rightarrow\infty$ the field goes to zero like $B \rightarrow {M\over 2 
a^2}$, whereas for $a\rightarrow 0$ the field tends to a non-zero value 
$B\rightarrow 1/(2M)$. In the second branch $B\rightarrow -2/M$ as 
$a\rightarrow\infty$. An analogous branch structure was found in \cite{gaha} 
for the Ernst metric, where the second branch was found to be somewhat 
anomalous. We will not discuss that here, and in the following we will only 
consider the first branch of solutions (upper signs) in \reef{bvalue}. Observe 
that values of $B$ larger than \reef{bvalue} would have yielded a cosmic string 
stretching along the dipole, a ``dumbbell" configuration similar to that 
considered in \cite{mypairs}. 

We have therefore succeeded in removing the conical singularities of the Bonnor 
dipole solution. However, the metric still becomes singular at the endpoints of 
the dipole, $(r=r_+, \theta=0)$ and $(r=r_+, \theta=\pi)$. Remarkably, we can 
show that these singularities are merely artifacts of the coordinate system. In 
order to do so, let us study the geometry of the region very close to these 
points. To this effect, change the coordinates $(r,\theta)$ to $(\rho,\bth)$ 
as\footnote{A similar change was performed in \cite{sen} in a study of the 
Kaluza--Klein dipole, which can be recovered as a particular case of solutions 
described below.} 
\beqa
r&=&r_+ +{\rho\over 2} (1+\cos\bth)\ ,\nonumber\\
\sin^2\theta &=&{1\over \sqrt{M^2+a^2}}\rho(1-\cos\bth)\ ,
\labell{change}
\eeqa
and take $\rho$ to be much smaller than any other length scale involved so as
to get near the poles. 
In this limit the solution becomes, near $(r=r_+, \theta=0)$,
\beq
ds^2 =-{\rho^2\over Q^2} dt^2 + {Q^2\over \rho^2} d\rho^2 +Q^2 (d\bth^2 
+\sin^2\bth d\varphi^2) \ ,
\labell{throat}
\eeq
\beq
A_\varphi=-Q (1-\cos\bth) \ ,
\labell{monopole}
\eeq
where
\beq
Q={ M r_+ \over \sqrt{M^2 +a^2}}\ .
\labell{charge}
\eeq
This is precisely the Bertotti--Robinson solution, AdS$_2\times S^2$, which 
describes the near horizon limit of an extreme Reissner--Nordstr\"{o}m black 
hole with charge $-Q$. In a similar way, at the other endpoint $(r=r_+, 
\theta=\pi)$ we find the same geometry but this time with the opposite 
charge\footnote{The signs of the charges would be reversed for the second 
branch in \reef{bvalue}.}. Therefore, the solution contains regular horizons at 
the poles, and it can be continued beyond $\rho=0$. 

Apparently, the field $B$ has no effect down the throat \reef{throat}. But, 
crucially, realize that in order to arrive at \reef{throat} the field $B$ is 
required to take precisely the value \reef{bvalue}. It is illuminating to see 
how things change for other values of $B$. If we keep $B$ arbitrary, then the 
limiting form of the solution near the poles is
\beq
ds^2 =g^2(\bth)\left[ -{\rho^2\over Q^2} dt^2 + {Q^2\over \rho^2} d\rho^2 +Q^2 
d\bth^2 \right]
+{Q^2\sin^2\bth\over g^2(\bth)} d\varphi^2 \ ,
\labell{deformthroat}
\eeq
with $Q$ as in \reef{charge} above, and where
\beq
g(\bth)={1\over 2}\left[1+\cos\bth +\left({a\over\sqrt{M^2+a^2}}+BQ\right)^2
(1-\cos\bth)\right]
\labell{deform}
\eeq
is a function such that $g(\bth)=1$ when the field $B$ is tuned to the value 
\reef{bvalue}. The important point is that in general the surface $\rho=0$ is 
still a horizon, albeit not one of spherical symmetry. Instead, the horizon is 
a prolate spheroid, which is further distorted by a conical defect at either 
pole. We want to stress that the horizons are present even for the case of the 
Bonnor dipole ($B=0$). As far as we know, this crucial feature of the Bonnor 
solution \reef{bonnorm} has gone unnoticed in all previous literature.

The gauge field near the horizon is also distorted from its monopolar form to
\beq
A_\varphi=-Q\left( {a\over\sqrt{M^2+a^2}}+BQ\right) 
{1-\cos\bth\over g(\bth)} \ .
\labell{deformpole}
\eeq
The physical magnetic charge of the hole can now be computed using
Gauss's law,
$\hat Q={1\over 4\pi}\int_{S^2}F$
($S^2$ is any topological sphere surrounding the charge) so, in general,
the actual physical
charge of the hole is not
$Q$, but rather
\beq
\hat Q={\triangle\varphi\over
4\pi}\left(A_\varphi
(\bth=\pi)-A_\varphi (\bth=0) \right)=
{\triangle\varphi\over 2\pi}{Q\over {a\over\sqrt{M^2+a^2}}+BQ} \ .
\labell{qhat}\eeq

The limiting geometries above were valid for arbitrary values of $a$, as long 
as we remain close enough to the pole. If, instead, we consider the limit of 
very large $a$, while keeping $r-r_+$ and $a\sin^2\theta$ finite, the solution 
\reef{diholem} tends to
\beq
ds^2 =- \left( 1+{Q\over \rho}\right)^{-2} dt^2 +\left( 1+{Q\over 
\rho}\right)^2 [d\rho^2 +\rho^2 (d\bth^2 +\sin^2\bth d\varphi^2)] \ ,
\labell{rn}
\eeq
with $Q\rightarrow M$, and $A_\varphi$ as in \reef{monopole}. We
recognize this as the extremal Reissner--Nordstr\"{o}m black hole. In
the limit $a\rightarrow\infty$ the magnetic field $B$ vanishes,
consistently with the interpretation that the poles are ``infinitely
apart" from each other and the force between them goes to zero.
Incidentally, note that if one wanted to consider an adiabatic process
where the two black holes held in equilibrium are taken apart, then the
magnetic field would obviously have to be adjusted at every moment in
such a way that the field precisely balances the forces for fixed values
of the charge of each hole \reef{qhat}.

So we conclude that our solution \reef{diholem}, \reef{diholea} indeed
describes a dihole.  In general, for finite values of $a$, the geometry of the
black holes is distorted from their asymptotically flat form \reef{rn}, but,
for the particular value of $B$ in \reef{bvalue}, the distortion becomes
inappreciable well down the throat, where we recover the near horizon geometry
\reef{throat}.  Moreover, the infinite proper distance along the dipole line
$r=r_+$ is now seen as a consequence of the infinite throat characteristic of
extremal Reissner--Nordstr\"{o}m black holes.

The dihole character of the dipole solution brings 
about some interesting consequences (now we restrict ourselves to the solution 
with $B$ given by the upper sign solution in \reef{bvalue}). There is a 
non--vanishing area associated 
to the horizon of each of the black holes, and therefore an entropy. This is 
easily obtained from \reef{throat} as $S={\cal A}_h/4= \pi Q^2$ for each hole.

In the limit of large separation between the holes we would expect a Newtonian
approximation to become reasonable.  Indeed, for large $a$, the magnetic field
$B$ exerts the right force, $T\approx Q B \approx Q^2/(2 a^2)$ to 
counterbalance the attraction between two particles at a distance of the order 
of $a$. 

As $a\rightarrow 0$, though, there is the peculiarity that the black holes 
in \reef{diholem} never appear to merge. For $a=0$ the solution is 
non--singular (outside the horizons), and is to be interpreted as the 
configuration of minimal separation between the black holes. It corresponds to 
a maximum value for $QB$, namely $(QB)_{\rm max}=1$.

We have mentioned that the mass of the dipole was, for Bonnor's asymptotically
flat solution, equal to $2M$.  The solution
\reef{diholem}, \reef{diholea} is instead asymptotic to the Melvin universe,
but it is still possible to compute its energy by taking the Melvin universe as
the reference background, following \cite{haho}.  The result is that the energy
is still equal to $2M$.  In the limit of large separation the mass of each
black hole \reef{rn} is $M$, so the total energy is the sum of the energies of
the separate black holes. Thus, for infinite separation the interaction energy 
vanishes, as could have been expected.

Now, at finite values of $a$ we would expect to find a non--vanishing 
interaction energy. Given that extremal black holes in isolation satisfy 
$M_{\rm bh}=Q$, we can estimate the interaction energy in the dipole as
\beq
E_{\rm int}= E_{\rm total}-2M_{\rm bh}=2M -2Q=-{2 M^2  \over \sqrt{M^2 +a^2}}
\ .
\eeq
This is negative, reflecting the attraction between the black holes. For fixed 
black hole charge $Q$, this energy is minimized when $a=0$. Notice that the 
value as $a\rightarrow 0$ is $E_{\rm int}=-2 M$. 

Now let us turn to the generalization of these diholes to theories with a 
dilaton field $\phi$. The action we consider is
\beq
I={1\over 16\pi G}\int d^4x \; \sqrt{-g}\left(R-2(\partial\phi)^2 
-e^{-2\alpha\phi} F^2\right)\ .
\eeq
For dilaton coupling $\alpha=0$ we will recover the results discussed above. 
The case of $\alpha^2=3$ corresponds to Kaluza--Klein theory, and in this case 
the solutions admit nice geometric interpretations. The Kaluza--Klein analog 
of the Bonnor dipole was identified in \cite{gp}. The introduction of the 
background magnetic field, together with a thorough analysis of the structure 
of the solutions and extensions to higher dimensions, was undertaken in 
\cite{dggh2}.

For arbitrary values of the dilaton coupling, the counterparts of the
Bonnor dipole were obtained in \cite{davged}. The conical singularity
along the axis was correctly identified there. Indeed, a straightforward
calculation like the one in \reef{conein} shows that the conical deficit
is present for arbitrary values of $\alpha$. More importantly, we will
find that the entire dihole structure reveals itself for any value of
$\alpha$, in a manner entirely analogous to the Einstein--Maxwell
dihole. This is at variance with the conclusions in \cite{davged}, an
issue we will return to below, after completing our analysis. 

It is a straightforward matter to take the dipole solutions in \cite{davged} 
and subject them to a dilatonic Harrison transformation \cite{dgkt}. The 
resulting metric is
\beq
ds^2=\Lambda^{2\over 1+\alpha^2} \left[-dt^2 + {\Sigma^{4\over 1+\alpha^2}\over 
\left(\Delta +(M^2 
+a^2)\sin^2\theta\right)^{3-\alpha^2\over 1+\alpha^2}} \left({dr^2\over 
\Delta}+d\theta^2 \right)\right]+{\Delta 
\sin^2\theta \over \Lambda^{2\over 1+\alpha^2}}d\varphi^2 \ ,
\labell{dilholem}
\eeq
the dilaton, $e^{-\phi}=\Lambda^{\alpha\over 1+\alpha^2}$, and the gauge 
potential,
\beq
A_\varphi=-{{2\over \sqrt{1+\alpha^2}} M r a+{1\over 2} B[(r^2-a^2)^2+\Delta 
a^2\sin^2\theta] \over 
\Lambda\Sigma}\sin^2\theta \ ,
\labell{dilholea}
\eeq
with 
\beq
\Lambda={ \Delta +a^2\sin^2\theta +2 \sqrt{1+\alpha^2}B M r a 
\sin^2\theta+{1+\alpha^2\over 4} B^2 
\sin^2\theta[(r^2-a^2)^2+\Delta a^2\sin^2\theta] \over \Sigma} \ .
\eeq
and $\Delta$ and $\Sigma$ still given by \reef{defdesi}.

The analysis of these solutions can be carried out in the same manner as above 
for the Einstein--Maxwell dihole, only in this case at the poles we find 
extremal dilatonic holes, the horizons being replaced by null
singularities. The solutions asymptote to the dilaton Melvin 
solutions of \cite{gima}. The value of the magnetic field that removes all 
conical singularities is
\beq
B={2\over \sqrt{1+\alpha^2}}{M\over (r_+ + a)^2}
\labell{dilbvalue}
\eeq
(a second branch also exists for these cases). The same coordinate change as in 
eq.~\reef{change} yields, for large $a$,
\beq
ds^2= - \left( 1+{Q\over \rho}\right)^{-{2\over 1+\alpha^2}} dt^2 +\left( 
1+{Q\over 
\rho}\right)^{2\over 1+\alpha^2} [d\rho^2 +\rho^2 (d\bth^2 +\sin^2\bth 
d\varphi^2)] \ ,
\labell{dilrn}
\eeq
with a monopole potential $A_\varphi$ of charge $Q/\sqrt{1+\alpha^2}$, and 
dilaton $e^{\phi}=\left( 1+{Q\over \rho}\right)^{\alpha\over 1+\alpha^2}$. 
These solutions are the extremal dilatonic holes of \cite{gima}.

In the same manner as we have done before in the absence of dilaton, 
we can also keep $a$ finite, but go to
small values of $\rho$. In this way we recover the geometry near the
(singular) horizon $\rho=0$ of the extreme dilatonic hole, with the
parameter $Q$ defined as in \reef{charge}, and some angular distortion
when $B$ takes values different from \reef{dilbvalue}.
This is,
\beq
ds^2\rightarrow g(\bth)^{2\over 1+\alpha^2}\left[-\left({\rho\over
Q}\right)^{2\over 1+\alpha^2}dt^2+\left({Q\over
\rho}\right)^{2\over
1+\alpha^2}\left(d\rho^2+\rho^2 d\bth^2\right)\right]+
\left({Q\over\rho}\right)^{2\over 1+\alpha^2}{\rho^2\sin^2\bth\over
g(\bth)^{2\over 1+\alpha^2}d\varphi^2}\ ,
\labell{nearpole}\eeq
now with 
\beq
g(\bth)={1\over 2}\left[1+\cos\bth
+\left({a\over\sqrt{M^2+a^2}}+BQ\sqrt{1+\alpha^2}\right)^2
(1-\cos\bth)\right]\ .
\eeq
In these coordinates it is easy to compute the scalar curvature near the
poles $r=r_+$, $\theta=0,\pi$ since these loci correspond to $\rho=0$.
One finds
\beq
R\rightarrow{f(\bth)\over \rho^{2\alpha^2\over 1+\alpha^2}}\ ,
\labell{scalarr}
\eeq
where $f(\bth)$ is a certain function which is regular for $0\leq
\bth\leq \pi$. We can see that the scalar curvature diverges at $\rho=0$
{\it except when $\alpha=0$}.
Thus $\rho=0$ (\ie $r=r_+,\theta=0,\pi$) is, for $\alpha\neq 0$, a real
singularity. But this is just the well-known null singularity of
extremal dilatonic holes. These holes do not have any
Bekenstein--Hawking entropy associated. The proper distance between the
holes for $\alpha\neq 0$ is finite, since in that case the proper
distance to the (singular) horizon of each hole \reef{dilrn} is known to
be finite. Besides, for values of $B$ other than \reef{dilbvalue} the
geometry around the singular horizon is angularly deformed in a manner
similar to \reef{deformthroat}. 

A very different interpretation of the geometry was proposed in
\cite{davged}, where it was
claimed that for $\alpha=1$ (and only for that value)
{\it regular non-extremal horizons} are present 
at the poles $r=r_+$, $\theta=0,\pi$. The
analysis of \cite{davged} was based on a study of certain two-dimensional
sections of the solution, in particular of the geometry of the
two-dimensional section given by
$r=r_+$, $\varphi={\rm constant}$. It
was pointed out in that paper
that the {\it intrinsic} curvature of this two-dimensional metric is
divergent at $\theta=0,\pi$ unless $\alpha=1$. However, such a restricted
two-dimensional study cannot be conclusive, if only for the fact that
singularities of a submanifold do not in general correspond to
singularities of the full manifold. The analysis is
indeed misleading: The full four-dimensional
structure of the solutions near the poles is manifested using
the coordinates ($t$, $\rho$, $\bth$, $\varphi$) as introduced in
\reef{change}, and then eq.\ \reef{scalarr} 
explictly shows that only $\alpha=0$ yields a
non-singular four-dimensional curvature at $\rho=0$. This is just as
expected from the fact that we have recovered the geometries near the
horizon of {\it extremal} charged dilatonic black holes, of which only
the pure Einstein-Maxwell case possesses a regular horizon. The standard
analysis of the structure near the horizon of the extremal
Reissner-Nordstrom black hole and its limiting Bertotti-Robinson
geometry can be equally well applied to \reef{throat}. In particular,
future directed geodesics can cross each of the horizons at each pole,
so at both poles we have future horizons. Since the geometry is symmetric
under time-reversal there are also past horizons. Also, the coordinates
can be extended in the standard manner beyond the horizons. This forms
the basis of our claim that the non-dilatonic solution describes a {\it
time-symmetric} configuration with two extremal black holes, each with a
future and a past horizon. This is obviously at variance with the claim
in \cite{davged} of a {\it non-extremal\/} white hole (a past horizon)
at one pole and a black hole (a future horizon) at the other pole for
$\alpha=1$ (and a singularity for $\alpha\neq 1$).
Actually, this time-asymmetric interpretation is another artifact of the
restriction to the two-dimensional submanifold mentioned above. We would
also like to stress that our
interpretation is in consonance with the one given in \cite{dggh2} for
the case $\alpha=\sqrt{3}$, and extends it in a natural way to other
values of $\alpha$.

Let us now discuss some generic aspects of the dihole solutions we have
constructed.  First of all, we have shown that the solutions of \cite{bonnor} 
and \cite{davged} are properly interpreted, for arbitrary values of the dilaton 
coupling, as diholes, with the holes being kept in equilibrium by strings or 
struts. The horizon of each hole is deformed by the field created by the other 
hole, as well as by the conical defect. We have found that an external field 
can be applied and tuned so as to balance the system and remove the conical 
singularities. In that case, the external field precisely cancels the field 
created by the other hole, with the effect that the distortion disappears and 
the horizon is spherically symmetric.

On the other hand, the conical defects that pull apart the holes in the Bonnor 
dihole can be made more physical by regarding them as the limit of 
self--gravitating vortices that end on the black holes. Therefore they add to 
the catalog of solutions describing cosmic strings ending on black holes 
\cite{bunch,mypairs}.

Another aspect to note is that the configurations are expected to be
unstable. On physical grounds it is clear that a slight deviation from
the equilibrium configuration should set the black holes either in
runaway motion away from each other, or collapsing onto one another. As
a matter of fact, the instability of the dihole is known to be present
for the $a=0$ solution in the Kaluza--Klein case $\alpha=\sqrt{3}$. In
that case the solution can be related to the Euclidean Schwarzschild
instanton, which is known to have an unstable mode \cite{gpy}.

Indeed, the instability of these solutions fits in nicely with the existence of
instantons describing the pair creation of black holes in an external field
\cite{paircr} or in the breaking of a cosmic string \cite{bunch}.  The diholes
are to be seen as the sphalerons sitting on top of the potential barrier, under
which the tunneling process takes place.  Thus, the dihole solutions in this
paper are closely linked to the C and Ernst type of solutions that describe
black holes accelerating apart \cite{kin,ernst,dgkt}.

We have mentioned as well that, when held in the external field, it does not
appear to be possible to bring the black holes close enough to make them merge.
Fixing the charge, there is an upper limiting value for the magnetic field,
which is approached as $a\rightarrow 0$ for the branch with $B>0$.  In this
limit the two-black hole structure still persists.  This might be seen as
providing support for the cosmic censorship conjecture:  the merging of the
Reissner--Nordstr{\"o}m black holes, which would imply the annihilation of
charge and possibly a change in the spatial topology, might have led to a naked
singularity.  Nevertheless, notice that for the Bonnor dihole kept in
equilibrium by cosmic strings, the black holes actually merge as $a\rightarrow
0$, and then form a singularity.  However, these ``tests" must be taken with a
pinch of salt, since given the instability of the solutions, a gedanken
experiment in which the black holes are slowly moved towards one another does
not appear to be physically realizable.  Related analyses of cosmic censorship
can be found in \cite{gaha}.

Several extensions of the work presented in this paper seem possible.  First of
all, it is clear that electric diholes can be constructed simply by dualizing
the magnetic field to an electric field.  More interesting are generalizations
to theories with a richer field content.  Dilaton black holes with coupling
$\alpha=0,1/\sqrt{3},1, \sqrt{3}$ are known to occur in the low energy
description of string and M-theory compactified down to four dimensions.  They
admit an interpretation in terms of branes intersecting in higher dimensions,
with all the longitudinal and relative transverse coordinates being
compactified.  As one example among many possible embeddings, the
Reissner--Nordstr\"{o}m black hole can be obtained as an intersection of \eg
four equally charged D3-branes \cite{laba}.  We can lift our solution
\reef{diholem} to ten dimensions by suitably adding flat dimensions, and then
interpret it as an intersecting brane-antibrane configuration.  A similar lift
can also be done for the solutions with the other three special values of 
$\alpha$.

Now, when the charges of the branes are not equal to one another, the four
dimensional black holes appear as solutions to theories with four $U(1)$ gauge
fields and three independent scalars \cite{cvyo}.  It is likely that dihole
solutions for these theories can be constructed.  Indeed, the existence of C 
and Ernst--type metrics in such
$U(1)^4$ theories, describing pairs of black holes accelerating apart
\cite{composite} strongly suggests that it should be possible to construct
their static dihole counterparts.\footnote{The term ``dihole" has
also been used in this context in \cite{tomas} to refer to a different type of
solutions.}

On the other hand, it is less clear how to obtain non-extreme diholes.
Also, one might speculate on the possibility that the dihole be held in
equilibrium not by an external magnetic field, but rather by the expansion
produced by a positive cosmological constant.  To our knowledge, such solutions
have not been constructed yet.

\section*{Acknowledgements}

Correspondence with Bert Janssen, Sudipta Mukherji and Makoto Natsuume is 
gratefully acknowledged.  Work supported by EPSRC through grant
GR/L38158 (UK), and by grant UPV 063.310-EB187/98 (Spain).

\end{document}